\documentclass[twocolumn,aps,prd,amsmath,amssymb]{revtex4}
\bibpunct{}{}{,}{s}{}{,}
\usepackage{graphicx}   

\def\p {\partial}

\def\be {\begin{equation}}

\def\ee  {\end{equation}}

\def\bea {\begin{eqnarray}}

\def\eea {\end{eqnarray}}

\def\nn {\nonumber}

\begin{document}

\title{Critical behaviour in quantum gravitational collapse}

\author{Viqar Husain}

\email{vhusain@unb.ca}

\affiliation{Department of Mathematics and Statistics \\University of New Brunswick \\Fredericton, NB Canada E3B5A3}

\begin{abstract}

We study the gravitational collapse of an inhomogeneous  scalar field  with quantum gravity corrections  associated with singularity avoidance. Numerical simulations indicate that  there is critical behaviour at the onset of black hole formation as in the classical theory, but  with the difference that black holes form with a mass gap. 

\bigskip

\end{abstract}

\maketitle

\section{Introduction}

One of the outstanding problems in theoretical physics is the
incomplete understanding at the quantum level of the formation,
and subsequent evolution of black holes in a quantum theory of
gravity. Although a subject of study for over three decades, it is
fair to say that, in spite of partial results in string theory and
loop quantum gravity, there is no widely accepted answer to many
of the puzzles of black hole physics. This is largely because
there has been no study of quantum dynamical collapse in these
approaches. Rather, progress has focused mainly on explanations of
the microscopic origin of the entropy of  static black holes from
state counting.\cite{stringbh,lqgbh} A four-dimensional spacetime picture of black hole
formation from matter collapse, and its subsequent evolution is
not available in any approach to quantum gravity at the present
time.

This paper describes  an attempt to address this problem in the
context of Hawking's original derivation of black hole radiation:
spherically symmetric gravity minimally coupled to a massless
scalar field. This is a non-linear 2d field theory describing the
coupled system of the metric and scalar field degrees of freedom.

Gravitational collapse in the classical theory in this model has
been carefully studied numerically \cite{chop,gundrev}, but its
full quantization has never been addressed. The classical results
are well known; the onset of black hole formation is characterized by 
a scaling law for black hole mass and a self-similar behavior of the
field variables. 

In semi-classical theory, Hawking's calculation \cite{hawk1} uses the eikonal
approximation for the wave equation in a mildly dynamical
background, where the dynamics centers on the surface of a star
undergoing collapse. The essential content of it is the extraction
of the phase of the ingoing mode from an outgoing solution of the
scalar wave equation as a classically collapsing star crosses its
Schwarzschild radius.  According to this calculation, black holes create
particles that  originate near the event horizon.  The approximation breaks down
at the late stages of evaporation, where quantum gravity effects
are expected to become important. 

It is likely that a complete understanding of quantum dynamics in
the gravity-scalar field system \cite{earlywork} will resolve at least some of the outstanding 
problems of black hole physics associated with the late stages of black hole 
evolution. This has been the motivation for recent  work on the development  of a  
quantum framework for studying this model \cite{hw-sing} and for a proposal 
of a picture of the evolution process that incorporates singularity 
avoidance \cite{ash-bojo-bh}. 

In the following we give a summary of the results of classical gravitational 
collapse of a scalar field in spherical symmetry, together with some details of 
numerical techniques. We then make a quantum 
gravity motivated modification of the equations of the model designed to take
into account ``singularity resolution.''  Numerical simulations  of these equations
indicate that there is critical behaviour at the onset of black hole formation
as in the classical theory, but that black holes form with a  mass gap dependent on the
fundamental discreteness scale  and initial data parameters.  

\section{Classical collapse}

The Einstein equations coupled to a massless minimally coupled scalar field  take the 
form 
\begin{equation}
R_{ab} = 8\pi \p_a\phi\p_b\phi
\end{equation}
In spherical symmetry the resulting equations may be written with the metric ansatz 
\begin{equation}
ds^2 = -f(r,t)^2 dt^2 + g(r,t)^2 dr^2 + r^2 d\Omega^2,
\end{equation}
with $\phi = \phi(r,t)$, or in double null coordinates as 
\begin{equation} 
ds^2 = -4 \alpha(u,v) dudv + r^2(u,v) d\Omega^2. 
\label{uvmetric}
\end{equation}
with $\phi=\phi(u,v)$. 
In either form these can be rewritten as  two first order evolution equations and one constraint 
equation.  Numerical studies have utilized both forms for the four-dimensional equations 
\cite{chop, garfuv}, and for the three-dimensional case with negative cosmological contant where 
collapse produces the Banados-Teitelboim-Zanelli (BTZ) black hole. \cite{chop-pret, ho} 

The main result is that  gravitational collapse at the threshold of black hole 
formation exhibits critical behaviour.  For  a parameter $a$ in the initial data for the scalar field, numerical simulations give a mass formula
\begin{equation}
M_{BH}  \sim (a-a_*)^\gamma
\label{masseqn}
\end{equation}
where $a_*$ is a critical value of $a$ above which black holes form. Both $a_*$ and the critical exponent $\gamma$ are numerically determined.  Furthermore the field variables exhibit a discrete self-similarity of the form
\begin{equation} 
\phi(r,t) = \phi(re^\Delta,te^\Delta)
\end{equation}
where the constant $\Delta$ is also numerically computed. 

In the parametrization in eqn. (\ref{uvmetric})  with $\alpha(u,v) := g(u,v)r'(u,v)$, where $\prime$ denotes the 
derivative with respect to $v$, the field equations in four dimensions may be written in the compact form 
 \cite{vg1,vg2}
\begin{eqnarray}
\dot{r} &=& -\frac{\bar{g}}{2} \label{evo1}\\
\dot{h} &=& {1\over 2r^2} (h-\phi) \left( gr - 4 \bar{g}\right) \label{evo2}
\end{eqnarray}  
where dot denotes partial derivative with respect to $u$, and  we have defined 
\begin{eqnarray}
h &=& \phi + \frac{1}{4}\ r\phi' ,\\
g &=& \exp\left[8\pi \int_u^v  \frac{1}{r} (h-\phi)^2 \ dv\right],\\
\bar{g} &=& \frac{1}{2} \int_u^v g\  dv
\end{eqnarray}

A numerical integration scheme for  these equations \cite{garfuv, ho,vg1,vg2} proceeds by using a ``space'' $v$ discretization 
 \begin{equation}
h(u,v) \rightarrow h_i(u), \ \ \ \ \ \ \  r(u,v) \rightarrow r_i(u)
\end{equation}
to obtain a set of coupled ODEs, where $i=0,\cdots N$ specifies the $v$ grid.  Initial data for these
two functions with, suitable boundary conditions, are prescribed on a constant $u=u_0$ initial slice, from which the functions $g$ and $\bar{g}$ are constructed using a Simpson's rule integration. Evolution in 
$u$ ``time'' is performed using the forth order Runge-Kutta method. 

For a physical picture, it is convenient to specify the initial scalar field configuration as 
$\phi(r(v,u_0), u_0)$, since the metric function $r$ is a dynamical variable in these coordinates.
We use the data 
\begin{eqnarray}
r(u_0, v) &=& v \\
\phi(u_0, r(u_0,v)) &=& a r^2  \exp\left[   - \left(\frac{r-r_0)}{\sigma}   \right)^2       \right].
\label{dataphi}
 \end{eqnarray}
The initial data parameters  are $a,\sigma,r_0$. 
\begin{figure}
\includegraphics[height=3.8in,width=3in,angle=-90]{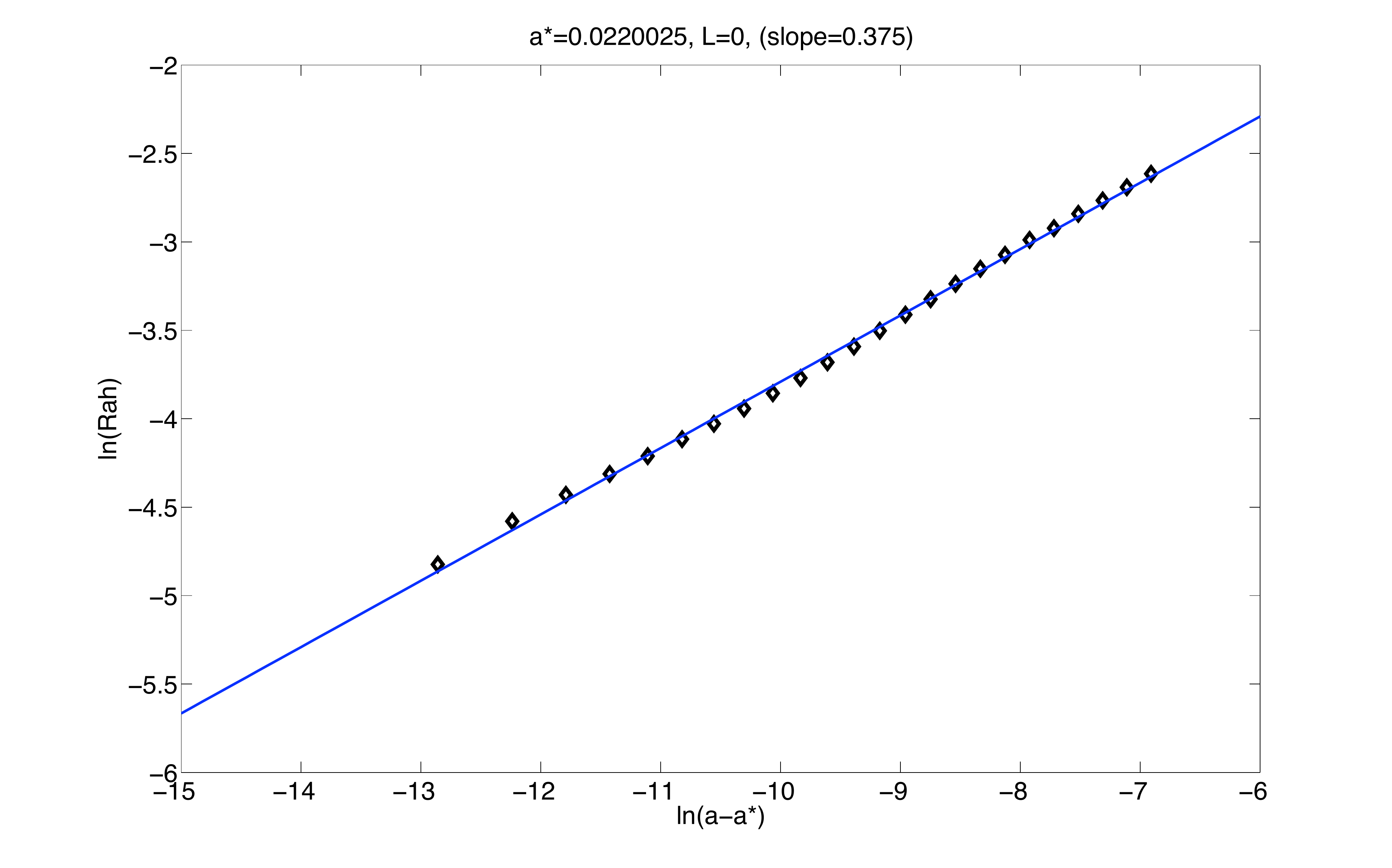}
\begin{center}
\caption{A log-log plot of black hole horizon radius as a function of initial data amplitude $a$ for vanishing 
fundamental discreteness scale $L$.  This is the expected classical result with slope $\gamma = 0.375$. The oscillation of points about the best fit line is a known feature of classical collapse.\cite{gundrev,hod-piran}}
\label{default}
\end{center}
\end{figure}

At each $u-$step of the evolution using Eqns. (\ref{evo1}-\ref{evo2}), the   function 
\begin{equation}
g^{ab}\partial_a r\partial_b r= - 2\frac {r\dot{r}}{g}
\end{equation} 
is computed. Its vanishing signals the formation of an apparent horizon. For each run of the code
with fixed amplitude $a$, this function is scanned from larger to smaller radial values after each
Runge-Kutta iteration. Evolution is terminated if a root is detected within a threshold value. The corresponding radial value is taken as the apparent horizon radius. This procedure is repeated
for a range of $a$ values until its  critical value $a^*$ is determined by a bisection method.  
For the subcritical case all radial grid points  evolve to zero without horizon detection.
This is the signal for complete reflection of the scalar field pulse.  

Figure 1 displays typical results of a run of such a code for the Gaussian scalar field profile specified above.  The results are in accord with other studies and serve as a useful check of our results.  
The slope of the line in this figure is $\gamma = 0.375$ giving a mass formula of the type (\ref{masseqn}) with the critical value $a^* =0.0220025$ (for $\sigma=1$ and $r_0=1$ in the initial data (\ref{dataphi}). The oscillation around the best-fit line is also a known feature of the critical behavior associated with classical collapse. \cite{gundrev}   This data is actually produced from a code for the quantum gravity corrected equations (to be described below), but  with the fundamental discreteness scale $L$ in these equations set to zero.  

We turn now to  describing an approach for  obtaining quantum gravity modifications of 
the above equations that incorporate singularity avoidance.

\section{Quantum Gravity Corrections} 

 The origin of quantum corrections  to gravitational collapse due to fundamental discreteness may be understood by considering quantum systems on a lattice. In spatial lattice based quantizations of field theory, local configuration variables are sampled discretely and momentum variables are realized  indirectly via translation operators. The lattice may be uniform (equispaced)  or irregular.  One can construct a non-separable Hilbert space whose basis elements are labeled by the lattice  chosen, in addition to other quantum numbers. Such a space may be viewed as the (infinite) sum of the Hilbert spaces, one for each lattice.

The kinematical Hilbert space of loop quantum gravity (LQG) is such a space. The classical variables used in LQG, the holonomy of a connection and a surface observable associated with its conjugate momentum,  are non-local. Their quantization is realized on a  space with basis states  labelled by graphs embedded in a three-manifold.  Operators may also be labeled by graphs, and hence connect states with different graph labels.  An example in LQG is a proposal for a ``graph changing'' hamiltonian constraint, \cite{QSDI}, which is defined to carry an edge that gets attached to a graph state.  Related quantizations are available for scalar field theory, \cite{polymer} and  a similar construction exists  for fermionic theories.  

\subsection{Quantum Mechanics on a Lattice}

 There are two types of   quantum gravity effects that can  arise from fundamental discreteness. These come from the way inverse configuration and momentum operators  must  be defined on a lattice. 
 Both are readily demonstrated by considering the mechanics of a particle on a lattice. 

A feature of spatial lattice theories is that there is a natural way to define inverse configuration operators via finite differencing. For a particle moving on the real line, a quantization on  a uniform lattice sampled from the real line with points 
$x_n = na$ ($n\in \cdots -1, 0 ,1 \cdots$) has an associated Hilbert space $|n\rangle$ with inner product 
\begin{equation}
\langle m|n\rangle = \delta_{mn}
\end{equation}
on which we have the operators 
\begin{eqnarray}
\hat{x}|n\rangle &= &na |n \rangle \\
\hat{U}_\lambda(p)|n\rangle&\equiv&\widehat{\exp(i\lambda p)}|n\rangle = |n-1\rangle.
\end{eqnarray}
We can define a densely defined operator corresponding to the variable $|x|^{-1}$, or other inverse configuration operators, by realizing a lattice finite difference scheme for the derivative 
of $f'(x_n)$ such as  
\begin{equation}
\frac{df(x_n)}{dx} \rightarrow \frac{f(x_n+\lambda) - f(x_n-\lambda)}{2\lambda}
\end{equation}
on the Hilbert space.\cite{jv}
For example applying it to the  function  $f(x) = \sqrt{|x|}$  gives the scheme 
\begin{align}
\frac{1}{|x|}
&=\left( 2\frac{d\sqrt{|x_n|}}{dx}\right)^2 \nn\\
& \rightarrow   \frac{1}{\lambda^2} \left(\sqrt{|x_n+\lambda|} - \sqrt{|x_n - \lambda|}\right)^2 
\label{finite-diff}
\end{align}
\begin{figure}
\hskip-0.4in 
\includegraphics[height=3.5in,width=4.0in,angle=0]{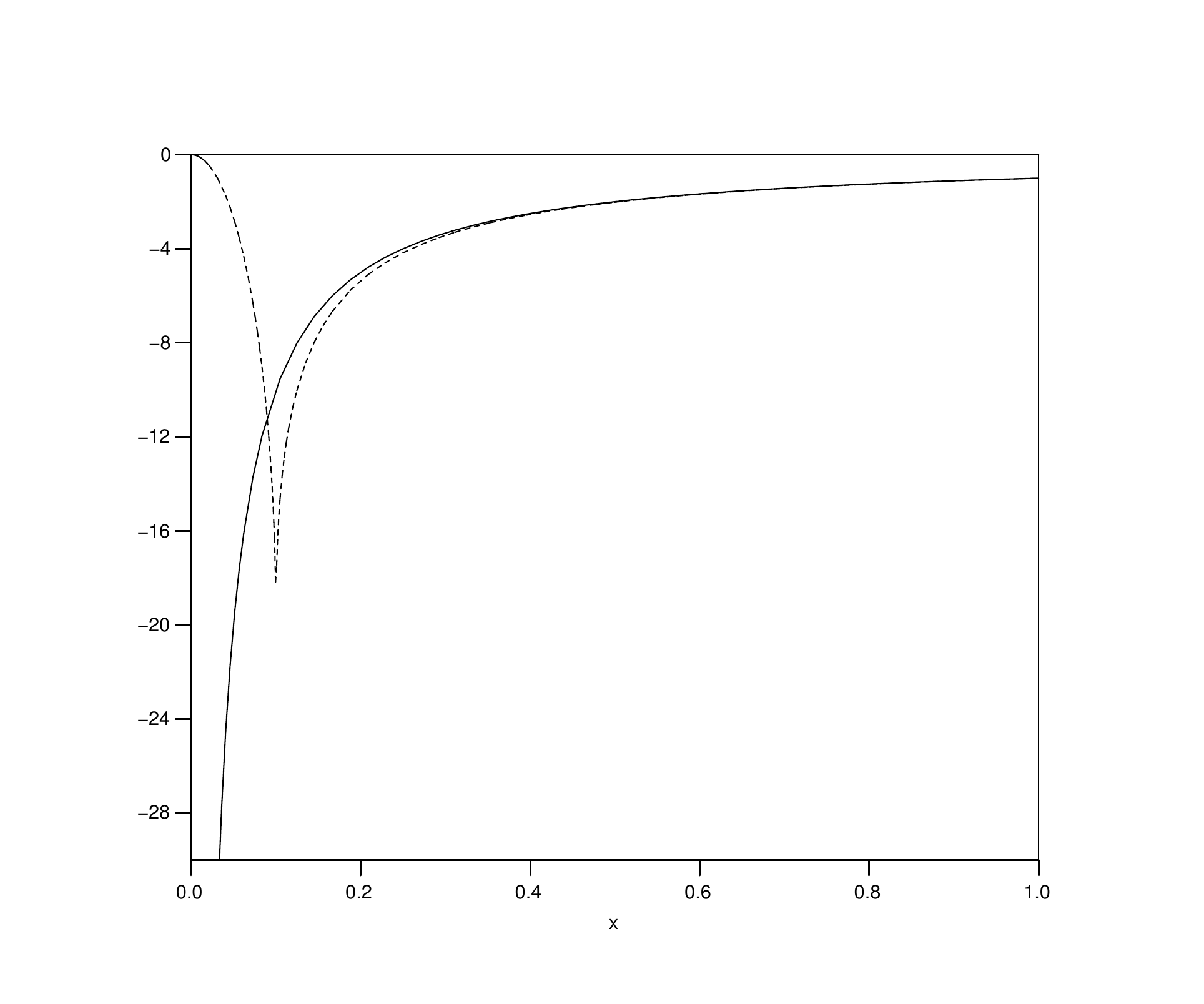}
\caption{The eigenvalue of the operator $-\widehat{1/x}$ Eqn.  (\ref{1/x-op}) (dotted line), and the function $f(x)=-1/x$ for $\lambda = 0.1$.}
\label{default}
\end{figure}    
The right hand side of this expression is the eigenvalue of the operator  
\begin{align}
\widehat{\frac{1}{|x|}}
&:=  \frac{1}{\lambda^2} 
\left( 
\hat{U}_\lambda \sqrt{|x|}\, \hat{U}_\lambda^\dagger
- 
\hat{U}_\lambda^\dagger \sqrt{|x|}\, \hat{U}_\lambda
\right)^2 \nn\\
&= \frac{1}{\lambda^2} 
\left( \hat{U}_\lambda^\dagger\left[\hat{U}_\lambda, \sqrt{|x|}\
\right] - \hat{U}_\lambda\left[\hat{U}^\dagger_\lambda, \sqrt{|x|}\
\right]\right)^2 , 
\label{1/x-op}
\end{align}
in the basis $|n\rangle$.  The eigenvalue  is bounded, and hence may be viewed as realizing an aspect of ``singularity avoidance'' in a theory with a fundamental discreteness scale.   

The finite difference idea captures the essence  of the definition on the inverse triad operator in LQG \cite{QSDI} and the inverse scale factor operator in loop quantum cosmology. \cite{martin1/a} In the general case, without symmetry reduction, such operators are not bounded unless these is restriction to  suitably defined semiclassical subspaces. \cite{brun-thie} Furthermore, it is clear that there are as many such operators as finite difference schemes in addition to the freedom in the choice of function $f(x)$; for instance a more general case is $f(x) = |x|^{1/k}$ for $k>1$. These are all examples of quantization  
ambiguities. 

So far we have seen an example, in a simplified context,  of an operator and its eigenvalue that arises from the necessity of defining operators  corresponding to inverse configuration variables. In quantum gravity  there is a need to define operators corresponding to inverse metric 
functions such as the determinant of the 3-metric, which arises in the hamiltonian constraint.  A second type of   modification  comes from the fact that momentum operators do not exist on a lattice. As already noted these must be defined indirectly using the translation operators $\hat{U}_\lambda(p)$.  One such definition is 
\begin{equation}
\hat{p}_\lambda := \frac{1}{i\lambda} \left( \hat{U}_\lambda - \hat{U}_\lambda^\dagger  \right) 
\end{equation}
 Such expressions lead to ``momentum corrections'' due to fundamental discreteness, which in quantum 
 gravity come from the variables canonically conjugate to the spatial metric -- the ADM momentum. The momentum operators can also be used to define lattice creation and annihilation 
  \be
A^\pm_\lambda  \equiv  \hat{x} \pm  i p_\lambda. 
 \ee  
   
In the following study of  gravitational collapse of a scalar field, we consider only the former type of correction, with the understanding that a complete treatment would require both.  The generalizations
of the operators to the field theory case are similar. 
 
\subsection{Gravity-Scalar Field Model} 

The application of this type of procedure  to the gravity-scalar field collapse problem in spherical symmetry requires a lattice-based Hamiltonian quantization scheme. This has been  developed recently.\cite{hw-sing} It contains a prescription for defining inverse field operators  to represent  variables such as  powers of $1/R(r,t)$, where $R(r,t)$ is the metric variable that is the measure of the size of spheres. ($r$ and $t$ are radial and time coordinates.)  Such factors arise in the Hamiltonian constraint and the field equations, as is evident from Eqn. (\ref{evo1}), (where we have referred to this function as $r(u,v)$). 

The quantization route we follow is unconventional in that field
momenta are not represented as self-adjoint operators; rather only
exponentials of momenta are realized on the Hilbert space. This is
similar to what happens in a lattice quantization, except that, as
we see below, every quantum state represents a lattice sampling of
field excitations, with all lattices allowed. Any numerical computation
is  of course only possible if it is restricted to a fixed lattice subspace.
In this sense non-separable  Hilbert spaces are not computationally
relevant.  

The gravitational phase space variables for the model are the configuration
variables $(R,\phi)$ and their conjugate momenta $(P_R, P_\phi)$.  
The basic Poisson bracket we consider is 
\bea
 \left\{R_f, e^{i\lambda P_R(r)}\right\}&\equiv& \left\{ \int_0^\infty Rf\ dr, e^{i\lambda P_R(r)} \right\}
\nn \\ 
 &=& i2G\lambda f(r) e^{i\lambda P_R(r)},
\label{bpb}
\eea
where $f$ is a suitable smearing function. 

This Poisson bracket may be represented  on a Hilbert space  with basis
\bea
&& |e^{i \sum_k a_k P_R(x_k)}, e^{i L^2\sum_l b_l
P_\phi(y_l)}\rangle \nn\\
&\ \ & \equiv |a_1\ldots a_{N_1};b_1\ldots
b_{N_2}\rangle, \label{basis}
\eea
where the factors of $L$ in the exponents reflect the length
dimensions of the respective field variables, and $a_k,b_l$ are
real numbers which represent the excitations of the scalar quantum
fields $R$ and $\phi$ at the radial locations $\{x_k\}$ and
$\{y_l\}$. The inner product on this basis is
\bea
&&\langle a_1 \ldots a_{N_1};b_1,\ldots b_{N_2}|a'_1 \ldots
 a'_{N_1}; b'_1\ldots b'_{N_2} \rangle \nn \\
 &\  \ & = \delta_{a_1,a_1'}\ldots \delta_{b_{N_2},b_{N_2}'},\nn
\eea
if the states contain the same number of sampled points, and is
zero otherwise.

The action of the basic operators are given by
\bea
&&\hat{R}_f\ |a_1 \ldots a_{N_1};b_1 \ldots b_{N_2}\rangle \nn\\
&\  \ &=
 L^2 \sum_k a_k f(x_k)|a_1 \ldots a_{N_1};b_1 \ldots b_{N_2}\rangle,
\eea
\bea 
&& \widehat{e^{i \lambda_j P_R(x_j)}}|a_1\ldots a_{N_1};b_1\ldots
b_{N_2}\rangle \nn \\
 &\ \ & = |a_1\ldots, a_j-\lambda_j,\ldots a_{N_1};b_1\ldots b_{N_2}\rangle,
\eea
where $a_j$ is $0$ if the point $x_j$ is not part of the original
basis state. In this case the action creates a new excitation at
the point $x_j$ with value $-\lambda_j$. These definitions give
the commutator
\be
\left[\hat{R}_f,\widehat{e^{i\lambda P_R(x)}} \right] =
-\lambda f(x) L^2 \widehat{e^{i\lambda P_R(x)}}.
\ee
Comparing this with (\ref{bpb}), and using the Poisson bracket
commutator correspondence $i\hbar\{\ ,\ \}\leftrightarrow [\ ,\ ]$
gives $L = \sqrt{2} l_P$, where $l_P$ is the Planck length. There
are similar operator definitions for the canonical pair
$(\phi,P_\phi)$.

This quantization is distinct from the LQG inspired ``polymer'' 
approach to field theory \cite{polymer}  where it is the configuration
variables that are represented in exponential form, following the 
representation of holonomy operators in LQG. The present approach is more 
akin to conventional  quantization in spatial lattice field theory.  It may be
viewed as the ``dual'' of the polymer approach.

\subsection{Inverse $R$ operators}

The  functionals  
\be 
R_f=\int_0^\infty dr fR 
\ee
used as the configuration variables serve as our starting point for defining 
inverse operators.     The procedure for doing this is similar
to that of the particle example.  Classical identities such as 
\be
\frac{1}{|R_f|} = \left(\frac{2}{i\lambda G f(r)}\
e^{-i\lambda P_R(r)}\left\{\sqrt{|R_f|}, e^{i\lambda P_R(r)} \right\}\right)^2,
\label{invRf}
\ee
where the functions $f$ do not have zeroes, may be used to define 
operators.  For example a suitable choice is a sharply peaked Gaussian 
at the point $r_k$, which also  serves to localize the operator. (It is useful to choose 
the Gaussians such that $f(r_k)=1$.)  

The representation for
the quantum theory described above is such that the operator
corresponding to $R_f$ has a zero eigenvalue. Therefore we
represent $1/R_f$ using the r.h.s. of (\ref{invRf}). The  
corresponding operator
\be
\widehat{\frac{1}{|R_f|}} \equiv
  \left( \frac{2}{l_P^2 \lambda f(x_j)}\widehat{e^{- i\lambda P_R(x_j)}}\left[ \widehat{\sqrt{|R_f|}},
\ \widehat{e^{i \lambda P_R(x_j)}} \right] \right)^2. \label{1/R}
\ee
 is densely defined and bounded.  This may be  illustrated   with the
basis state
\be
|a_0\rangle \equiv |e^{ia_0P_R(r=0)}\rangle,
\ee
which represents an excitation $a_0$ of the quantum field
$\hat{R}_f$ at the coordinate origin:
\bea
\hat{R}_f | a_0\rangle &=& (2 l_P^2) f(0) a_0\ |a_0\rangle, \\
\widehat{\frac{1}{|R_f|}}|a_0\rangle  &=& \frac{2}{(\lambda l_P
f(0))^2}\left(|a_0|^{1/2} - |a_0-1|^{1/2}\right)^2 |a_0\rangle  \nn
\eea
which is clearly bounded.  If there is no
excitation of $R_f$ at the coordinate origin, i.e. $a_0=0$, the
upper bound on the eigenvalue of the inverse operator is
$2/\lambda^2 l_P^2$. (In the units we are using, $P_R$ is dimensionless,
so $\lambda$ is a  dimensionless number, which can be taken to be unity.) 

A symmetrical version of this operator is defined as for the particle case 
in Eqn. (\ref{1/x-op}). Its eigenvalue on the basis state above is of the 
same functional form as in Eqn. (\ref{finite-diff})
\be 
\frac{1}{(\lambda l_P  f(0))^2} \left(\sqrt{|a_0+\lambda|} - \sqrt{|a_0 - \lambda|}\right)^2
\ee

\subsection{Semiclassical states}

The basis states and their associated eigenvalues are not the ones we consider 
useful for quantum corrections to the collapse problem discussed in Sec. 2. Rather 
we would like to find suitable semiclassical states  
\be
|R(r_k)\rangle_{sc}, 
\ee
where $R(r_k)$ is the sampling on a  uniform radial lattice of any prescribed function $R(r)$, with the 
properties that  for each lattice point $r_k$ we have 
\be
 \left\langle  \hat{R}_f   \right\rangle_{sc} = R(r_k) 
 \ee
 and
 \bea
\left\langle  \widehat{\frac{1}{|R_f|}}   \right\rangle_{sc} =  
&& \frac{1}{(\lambda l_P  f(r_k))^2}  \nn \\
&&\times   \left(\sqrt{|R(r_k)+\lambda|} - \sqrt{|R(r_k) - \lambda|}\right)^2
\label{1/Rexpec}
\eea
The right hand side of the latter function has the property that for $R(r_k)>> \lambda$, 
it behaves like $1/R(r_k)$ but has a different functional form
for $R(r_k)\sim \lambda$, just as in the particle mechanics case. 

It is possible to explicitly construct such states.\cite{vh-dt} The idea is to associate 
semiclassical states for a single particle on a line \cite{hw-semi} with the lattice points   $\{r_k\}$, 
such that  a state at point $r_i$ is peaked at coordinate value $R(r_i)$. 
The state $|R(r_k)\rangle$ is then defined to be the product of such single particle states. 
It represents a quantum state corresponding to the classical profile $R(r)$.
(It is convenient to take $f(r_k)=1$ and set $\lambda l_p = L$, which we do in the following. )

\subsection{Modified Collapse Equations }

A derivation of quantum gravity corrections to the field equations requires a number of inputs depending on the approach taken.  One approach is to make use of semiclassical states peaked on classical configurations.    Given a definition  of constraint operators on a kinematical Hilbert space, one then computes expectation values of the constraints in such states. This results in quantum corrected ``effective constraints,'' which to leading order in $\hbar$ are the classical constraints. The idea is to use these new constraints to derive evolution equations. For example, if  $\hat{H}(\hat{x},\hat{p})$ is such an operator, its expectation value in a semiclassical state $| \tilde{x}, \tilde{p}\rangle$ would give  a function $H(\tilde{x},\tilde{p})$.  This function would then be used to derive quantum corrected equations.  This is the approach we take.  Similar methods has been used to derive effective equations in loop quantum cosmology,\cite{bojowald-effec-cos}  and partly form the motivation for our work.  

It is apparent from sample computations that this approach gives the classical equations  in suitable limits. For example if there is a $1/x$ factor in a function $H(x,p)$ (as is the case in the Hamiltonian constraint), the equation of motion also contains a power of $1/x$. The corresponding quantum corrected equation will have a power of the spectrum function of the corresponding operator as in Eqn. (\ref{finite-diff}).  Given the form of this function, it is apparent that the large $x$ limit gives the classical equation as shown in Figure 2, and that the small $x$ region represents a ``repulsion,'' if the eigenvalue
represents a potential energy. 

A possible (and perhaps obvious) criticism of this approach to obtaining quantum corrected equations   is that one is deviating from the ideal of solving the quantum constraints a'la Dirac and  obtaining the physical Hilbert space before proceeding  to study physical  questions.  This approach has however not yielded any results beyond simple mini-superspace reductions, in the metric or the connection  variable constructions.  In order to proceed it is important to see what quantum gravity effects can be extracted from a kinematical construction of constraint operators.  The approach outlined above is one example of such a procedure, but it is an approximation which requires careful scrutiny especially with regard to
 the consistency of the quantum corrected constraints to the appropriate order.\cite{bojowald-effec}    
 
The eigenvalue of the $\widehat{1/R}$ field operator on basis states contains the modulus of the 
eigenvalue of the radial field operator $\hat{R}$. This causes a numerical problem in the quantum
modified equations where it is necessary to compute  derivatives with respect to a radial coordinate. 
To avoid this issue  we replace the expectation value (\ref{1/Rexpec}) by the smoothed  version 
\be 
\left\langle \widehat{\frac{1}{R_f}}\right\rangle \rightarrow  \frac{1}{R}\  \left(1- e^{-(R/L)^2}\right)  
\label{smooth1/R}
\ee
for numerical calculations. This form has the same qualitative features as the actual eigenvalue,
i.e. the large $R$ behaviour is $1/R$ with repulsion at small $R$, as in Fig. 2.  It is possible that
an exact computation of this expectation value in semiclassical states will provide a smooth function, 
so that such an approximation would become unnecessary. However we expect that the qualitative features of the result described below will remain unchanged. 

Modified collapse equations may be obtained by first computing effective constraints and obtaining equations of motion from them, or by directly replacing the expectation value  (\ref{smooth1/R}) in place of factors of $1/r$ into the equations from Sec. II. These two procedures of course give different effective equations. The procedure we used amounts to the latter since all the corrections from the former approach have not been incorporated into the code used.  The modification is such that  for large spheres (i.e. large $R$), the equations converge to the classical equations, with quantum gravity corrections  confined to smaller $R$ values determined by the fundamental discreteness scale $L$. As already noted, we do not include the momentum corrections in the present work; this would amount to replacing momenta in the constraints such as $P_R$ by the compactified form  $\sin(\lambda P_R)/\lambda$.\cite{hw-semi}
\section{Results}
 
The numerical procedure used for the quantum gravity corrected equations is identical to the
one outlined in Sec. II.   The simulations were performed for a range of $L$ values, with the 
integration lattice chosen such that time and space steps ranged from $10^{-3}$ to $10^{-4}$.
The initial data used was the same as that for the classical case. 
 
The results for horizon radius $R_{bh}$ as a function of the initial data parameter $a$ are given in 
Fig.  3, for fundamental discreteness scale $L$ values ranging from 0 to 0.2. There are a number of 
points to note here: (i) $L=0$ gives no mass gap, and the expected classical result. (ii)
Mass gaps at the onset of black hole formation are evident, with the critical value of the amplitude  
$a^*$ dependent on $L$;  the gaps increase with $L$. (iv) For amplitudes $a >> a^*$ the points 
corresponding to different $L$ values begin to merge, i.e.  for sufficiently massive initial 
data, the black hole masses become independent of $L$.  This is expected intuitively because 
bigger masses  should be less  affected by smaller scale quantum effects than smaller masses. 
(v) The critical amplitude $a^*$ depends on $L$ in a rather unusual way: for non-zero $L$, $a^*$ decreases with increasing $L$.  This means that black holes form more readily for larger $L$ but
the effect is rather small as far as the parameter ranges are concerned.  

Our results in Fig. 3 can be summarized in the black hole mass formula 
\be 
M_{BH} = m_0(L,a) + k\left[a-a^*(L)\right]^{\gamma(L,a)},
\ee
in the supercritical region $a>a^*$, where $m_0$ is the mass gap and $k$ and $\gamma$
are numerically determined constants.  It is apparent from the data that  the mass gap $m_0$
increases with $L$, and that the exponent $\gamma$ is also a mild function of $L$. Dependance
on the amplitude $a$ is also apparent in that for large values, the points appear to converge.

\begin{figure}
\includegraphics[height=3.8in,width=3in,angle=-90]{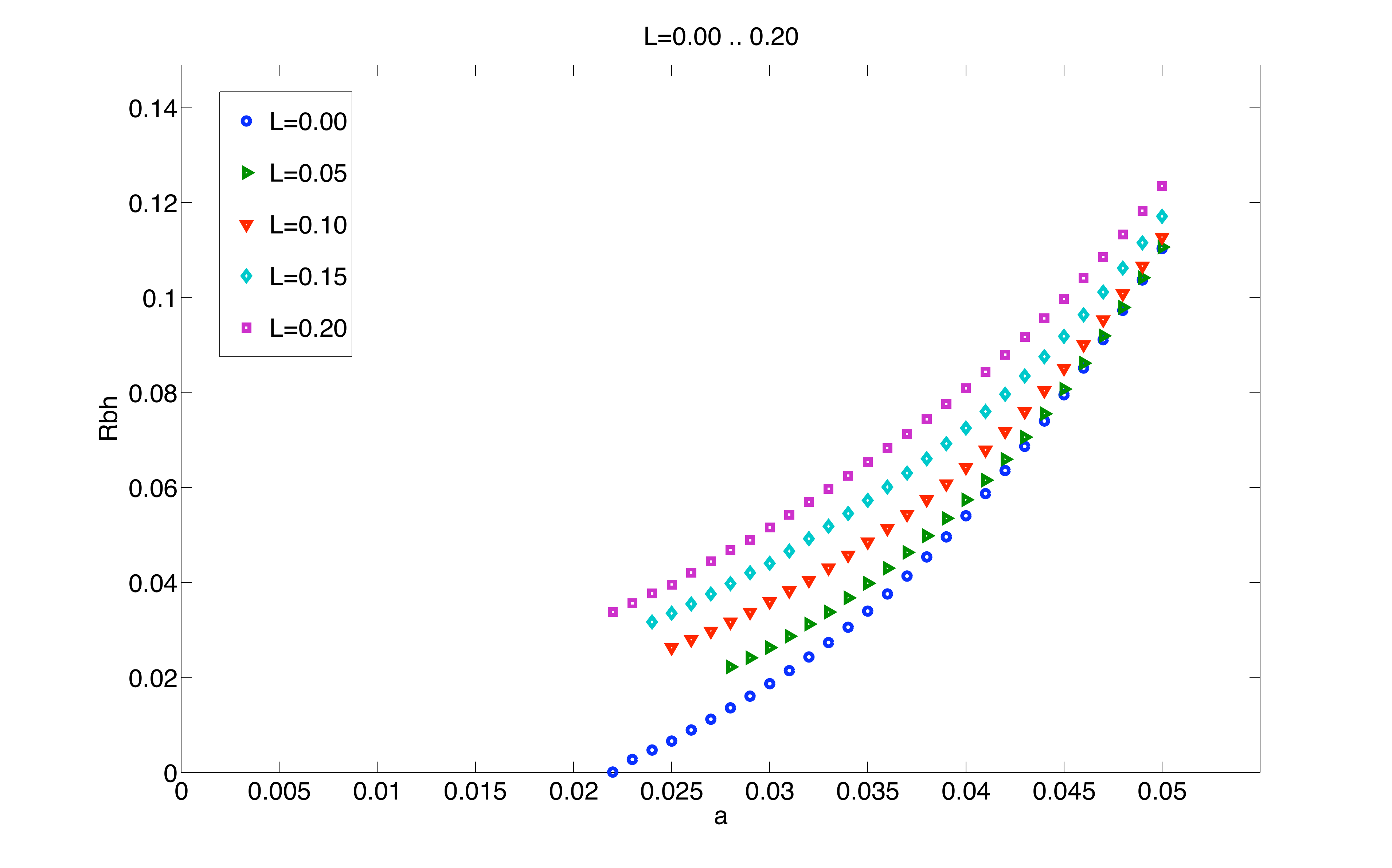}
\begin{center}
\caption{Black hole radius as a function of initial data amplitude $a$ for various values of the
fundamental discreteness scale $L$.  Black hole mass gap increases with $L$. }
\label{default}
\end{center}
\end{figure}

\section{Summary and Discussion}

We have described a first exploration of possible quantum gravity corrections in the gravitational collapse of an inhomogeneous scalar field in spherical symmetry. Numerical simulations indicate that there is critical behavior at the onset of black hole formation, and that black holes form with a mass gap unlike in  classical theory. 
  
There are general grounds to expect the mass gap result given that there is a fundamental 
discreteness scale in quantum gravity. This is simply that any physical object  
must have mass and size in units of this scale. It is gratifying that this is borne out, 
especially in the details of its dependence on both this scale and on the amplitude of the initial 
data. 

A mass gap in black hole formation has been noted before in a model where an exterior 
generalized Vaidya solution is patched to an interior Friedmann-Robertson-Walker model \cite{lqcgap}. 
This is also related to singularity avoidance in that the interior incorporates the "inverse scale factor"   
 bound used in  quantum cosmology.\cite{martin1/a, hw1/a} We note that there is no critical behaviour
 in this model because there is only matter inflow, and a black hole always forms by construction regardless of the initial data. This of course is a feature of Oppenheimer-Snyder type models of
 gravitational collapse with pure inflow. Only models with both matter inflow and outflow (i.e. metric functions fully  dependent  on both $r,t$ or $u,v$) have the possibility of critical behaviour,  with long time evolution leading either to black hole formation or full reflection.  
 
There are a number of directions for further work based on the approach we have used. 
These include incorporating the momentum modification into the evolution equations,
and a larger exploration of  the parameter space. With this in place it would be especially interesting to explore the nature of the exactly critical solution $a=a^*$; in the classical theory this is a naked singularity that realizes a finely tuned violation of the cosmic censorship hypothesis. With quantum gravity corrections the critical solution might be a  boson star.\cite{bosonstar}
  
A further challenge is to set up a numerical procedure that can be used to evolve data 
past horizon formation. This  will likely require coordinates such as the flat slice ones \cite{hoflat}, that have the potential to reveal how the scalar field and  horizon evolve beyond its formation. If the horizon is found to shrink, it would be accompanied by a flux of scalar field away from the 
hole region, and  should leave a clear numerical signature in the evolution of this field.
 An application of our approach to the possibly simpler $3-$dimensional case with negative
 cosmological constant may be easier to study in this respect.  
 
 Another direction with regard to such questions is the path integral. There is a derivation
of black hole  radiance in this approach \cite{pathint-rad}, and it would be interesting to see
how quantum gravity corrections arise in the spin foam \cite{spinfoam} approach.  Given the understanding of $3-$dimensional gravity \cite{3dfoam} using spin foam methods, this may be a first example to consider.

\medskip

\noindent{ \bf Acknowledgments} 
  
This work was supported by the Natural Science and Engineering Research Council of Canada.
I would like to thank   Bei-Lok Hu and Gabor Kunstatter for discussions.

\end{document}